\documentclass{article}
\title{General Relativistic Models for the Electron}
\author{S. M. Blinder}
\date{ }
\begin{document}
\newcount\eq
\def\eqn{\eqno(\the\eq)}
\def\ad{\advance\eq by1}

\centerline{\bf General Relativistic Models for the Electron}
\bigskip
\centerline{S. M. BLINDER}
\centerline{University of Michigan}
\centerline{Ann Arbor, MI 48109-1055 USA}
\centerline{(sblinder@umich.edu)}
\bigskip 

\begin{abstract}

{\noindent A model is proposed for the classical electron as a point
charge with finite electromagnetic self-energy. Modifications of the
Reissner-Nord\-str\o m (spin 0) and Kerr-Newman (spin $1\over 2$) solutions
of the Einstein-Maxwell equations are derived. It is conjectured that spacetime
curvature very close to the point charge deforms the electric and magnetic
fields such as to reduce the self-energy to a finite value by means of
Hawking polarization of the vacuum, much like that around a black
hole.}
  
\end{abstract}
\bigskip

\section{Introduction} 

In a recent paper\cite{1}, we showed
how a phenomenological representation of vacuum polarization could be
introduced into the formalism of classical electrodynamics. Modification of the
classical theory is significant only within the classical radius of the electron ($
r_0 = e^2/mc^2
\approx 2.8\times 10^{-13}$ cm) but this makes possible a consistent model
for point charges with finite electromagnetic self-energies. Further, the
self-interaction of a charged particle with its own electromagnetic field was
shown to be equivalent to its reaction to the vacuum polarization. 

In this paper, we reformulate these results in the context of general relativity.
We obtain two new solutions of the Einstein-Maxwell equations for a point
charge by modification of the Reissner-Nordstr\o m  and Kerr-Newman
metrics.  In contrast to the originals, our solutions possess {\it finite}
electromagnetic self-energies. The stress-energy tensors in our solutions imply
the presence of  finite charge distributions. These are conjectured to represent
vacuum polarization produced by the ultrastrong electromagnetic fields in the
vicinity of point charges, analogous to the phenomena proposed by
Hawking\cite{2} for strong {\it gravitational} fields. 

It was shown in [1] that, for the electron treated as a point charge and point mass,
an effective dielectric constant for the surrounding vacuum 
\ad$$\epsilon(r)=\exp(r_0/2r)\eqn$$\ad
leads to a self-energy entirely electromagnetic in origin, viz
$$W={1\over{8\pi}}\,\int\,{\bf E\cdot D}\, d^3{\bf r}=
{1\over{8\pi}}\,\int_0^\infty \, {e^2\over{\epsilon(r)\, r^4}}\, 4\pi r^2 \, dr
=mc^2         
\eqn$$\ad The net charge density (polarization plus free) around an isolated
electron is accordingly given by
$$\varrho(r)={e r_0\over{8\pi r^4}}\,\exp(-r_0/2r) \eqn$$\ad

Previous applications of general relativity to the structure of the
electron have  usually been based on perfect fluid
models involving additional non-electro\-magnetic forces\cite{3}-\cite{5}.  
A number of other workers have considered elementary particles from the viewpoint of
general relativity\cite{6}-\cite{10}. In some instances {\it negative}
gravitational mass was introduced in the interior of the particle in
order to achieve a finite rest energy. We find such assumptions unappealing. In this
paper, we propose a model for the electron based on the following premises: 1.  The
electron is a point charge; 2.  The electromagnetic self-energy is finite and
accounts for the total rest mass. We thus reject any non-electromagnetic
contributions to the electron mass. This is consistent with the nearly zero rest mass
of the electron's uncharged weak isodoublet partner---the neutrino; 3. Very close to
the electron, for distances
$r
\ll  r_0$, the electromagnetic energy density becomes sufficiently massive to fall
within the scope of general relativity.  Specifically, the curvature of spacetime
deforms the electromagnetic fields sufficiently to reduce the self-energy to a
finite value. 

\medskip
\section{Scalar Point Charge} We consider first the hypothetical electron
without spin or magnetic moment, described by a static,
spherically-symmetrical metric of the form
$$ds^2=-f(r)\,dt^2+{1\over f(r)}\,dr^2+r^2\,d\theta^2+r^2 \sin^2\theta\,d\phi^2
\eqn$$\ad The Reissner-Nordstr\o m metric\cite{11,12} has this structure with
$$f_{\scriptscriptstyle{\rm RN}}(r)=1-{2M\over r}+{Q^2\over r^2}\eqn$$\ad
where $M$ and
$Q$ are mass and charge in geometrized units:
$$M=Gm/c^2,\qquad Q=\sqrt{G}e/c^2$$  Solution of the Einstein-Maxwell
equations
$$R_{\lambda\mu}-{1\over2}R\,g_{\lambda\mu}={{8\pi G}\over{c^4}}
\,T_{\lambda\mu}\eqn$$\ad gives the stress-energy tensor
$$T^0_0=T^1_1=-T^2_2=-T^3_3\,=\,-{e^2\over 8\pi r^4}\eqn$$\ad This
represents a point charge since
$$ 4\pi T^\lambda_\mu=F^{\lambda\nu} F_{\mu\nu}-{1\over4}
g^\lambda_\mu  F^{\nu\sigma} F_{\nu\sigma}\eqn$$\ad implies that the only
non-zero field tensor components are
$$F_{10}=-F_{01}=E_r=e/r^2\eqn$$\ad The electromagnetic self-energy, given
by
$$W=-\int_0^\infty\,T^0_0\,4\pi r^2\,dr\eqn$$\ad remains divergent. For the
case $Q>M$, which obtains here, $g_{00}$ in the metric (4) has no real roots,
indicating the absence of a horizon. The electron is {\it not} a conventional
black hole, despite the singularity at $r=0$.

To obtain the desired electromagnetic self energy $mc^2$, we do some reverse
engineering on the Reissner-Nordstr\o m (RN) metric, replacing the function
(5) by
$$f(r)=1-{2M\over r}\exp(-Q^2/2Mr) \eqn$$\ad This is actually a very small
change, differing from the original form by less than 1 percent down to the
Planck length ($ \sqrt{\hbar G/c^3}\approx 1.6\times 10^{-33}\,{\rm cm}$).
With removal of the singularity at
$r=0$, the metric now becomes Lorentz flat  as $r\to 0$ (as well as $r\to
\infty$).  Note that
$$Q^2/2Mr=e^2/2mc^2r=r_0/2r \eqn$$\ad where $r_0$ is the classical
electron radius. The modified metric in the Ein\-stein-Maxwell equations can
be solved for a stress-energy tensor with the nonvanishing components:
$$T^0_0=T^1_1=-{e^2\,e^{-r_0/2r}\over 8\pi r^4},\qquad
T^2_2=T^3_3={e^2\,e^{-r_0/2r}\over 8\pi r^4}(1-{r_0\over 4r}) \eqn$$\ad  The
electromagnetic energy is now given by 
$$W=-\int_0^\infty\,T^0_0\,4\pi r^2\,dr=\int_0^\infty\,{e^2\,e^{-r_0/2r}\over
8\pi r^4}\, 4\pi r^2\,dr={e^2\over r_0}=mc^2
\eqn$$\ad  As advertised, we obtain a finite electron rest mass, entirely
electromagnetic in origin, the result originally sought by Lorentz. 

The Ricci scalar for our metric is nonvanishing:
$$R={Q^4\over 2 M r^5}\,e^{-Q^2/2Mr} \eqn$$\ad impling a 
gaussian curvature of spacetime absent in the original RN solution, which is
Ricci flat. 

The original Reissner-Nordstr\o m model can be interpreted as a point charge
in a hypothetical ``bare" vacuum. The modified metric implies some sort of
finite electron charge distribution. We conjecture that this represents
quantum-electrodynamic vacuum polarization.  The field tensor
$G_{\lambda\mu}$ for a polarizable medium is obtained from
$F_{\lambda\mu}$ by the substitutions
$E\to D$ and $B\to H$, such that (8) generalizes to Minkowski's stress-energy
tensor:
$$ 4\pi T^\lambda_\mu=F^{\lambda\nu} G_{\mu\nu}-{1\over4}
g^\lambda_\mu  F^{\nu\sigma} G_{\nu\sigma} \eqn$$\ad In particular,
$$ T^0_0= -{1\over 8\pi}\, E_r D_r \eqn$$\ad The stress-energy tensor (13) is
consistent with
$$D_r={e\over r^2},\qquad E_r={e\over r^2}\,e^{-r_0/2r} \eqn$$\ad  as
already anticipated in Eq (1). The electric displacement $D$ corresponds to the
Reissner-Nordstr\o m field from the ``free" electron point charge, while the
electric field $E$ takes account of all charges including those induced in the
vacuum. In this way we preserve the point-charge structure of the bare
electron, namely the validity of Gauss' law
$$\int\,{\bf D(r)\cdot dS} ={e\over r^2}\,4\pi r^2 = 4 \pi e \eqn$$\ad  for
arbitrarily small radius $r$. The introduction of an inhomogeneous dielectric
constant
$\epsilon(r)=\exp(r_0/2r)$  follows a suggestion by Weisskopf\cite{13}. As in
the case of a dense plasma, the dielectric constant increases with charge
density. In this case, the ``plasma" is made of virtual positron-electron pairs
produced by vacuum polarization\cite{14}. 

The nonvanishing trace of the stress-energy tensor (13) indicates
contributions other than the electromagnetic-field. One might interpret the
additional terms in $T^2_2$ and
$T^3_3$ as the presence of a viscous fluid, with components $p_\theta$ and
$p_\phi$, representing tangential pressure exerted by the virtual particles. 
Accordingly, the fields (18) do {\it not} represent a solution of
Maxwell's equations, except in the limit $r_0 \to 0$.

\medskip
\section{Electron with Spin}
\noindent  We extend our model to include electron spin by drawing from the
theory of rotating black holes.  Kerr\cite{15} first solved Einstein's equations
for a black hole with angular momentum.  Newman\cite{16} generalized this
result to include electrical charge.  The metric in Kerr-Newman geometry, in
the coordinates introduced by Boyer and Lindquist\cite{17}, can be written

$$\displaylines{ds^2 =
-{\Delta\over\rho^2}\,\big[dt-a\,\sin^2\theta\,d\phi\big]^2\hfill\cr\hfill
+{\sin^2\theta\over\rho^2}\,\big[(r^2+a^2)d\phi-a\,dt\big]^2
+{\rho^2\over\Delta}\,dr^2 +\rho^2\,d\theta^2 \qquad(20)\cr}$$\ad where
$$\rho^2\equiv r^2+a^2\cos^2\theta\eqn$$\ad and
$$\Delta\equiv a^2 + r^2 f(r) $$ The spin angular momentum per unit mass is
represented by the parameter
$$a\equiv S/M $$ For the original Kerr-Newman metric, $f(r)$ is given by Eq
(5), and 
$$\Delta_{\scriptscriptstyle\rm KN} = r^2 -2Mr +a^2 +Q^2$$   Computation of the
electromagnetic self energy $W$ with this metric gives a divergent result,  just
as in the Reissner-Nordstr\o m case.

It is therefore suggested that we try the modified functional form for $f(r)$ in
Eq (11). Following are explicit expressions for the nonvanishing elements of
the metric tensor (20):

$$g_{00}=-1-{r^2[f(r)-1]\over \rho^2} \qquad\qquad 
g_{11}={\rho^2\over{a^2+r^2f(r)}}$$
$$g_{22}=\rho^2 \qquad\qquad g_{33}=(a^2+r^2)\sin^2\theta- {a^2r^2
[f(r)-1]\over
\rho^2}\sin^4\theta$$
$$g_{30}=g_{03}={a r^2 [f(r)-1]\over \rho^2} \sin^2\theta\eqn$$\ad The
corresponding Jacobian is
$$\sqrt{-g}=\rho^2\sin\theta\eqn$$\ad independent of $f(r)$.

The stress-energy tensor follows after a lengthy computation. We display only
the $0\atop 0$ component:
$$\displaylines{T_0^0={{c^4/G}\over{16\pi\rho^6}}
\bigg[\big(-4a^2 r^2 - 4 r^4 + 2 a^2 \rho^2 +  4 r^2 \rho^2 - 2 \rho^4 \big)
\cr  +\,\big(4 a^2 r^2 +4 r^4 -2a^2 \rho^2-4 r^2 \rho^2+  2 \rho^4\big)
f(r) \cr       +\,\big(4 a^2 r^3 + 4 r^5 - 4 {a^2} r \rho^2 - 6 r^3 \rho^2 +   4 r
\rho^4\big)f'(r) \cr\hfill +\,\big(- a^2 r^2 \rho^2   - r^4 \rho^2 + r^2 \rho^4
\big) f''(r)\bigg]
\qquad (24) \cr}$$\ad with $\rho$ defined in Eq (21).

The electromagnetic energy is given by
$$W=-\int\,T_0^0\,\sqrt{-g}\,dr\,d\theta\,d\phi \eqn$$\ad Integrating over
angles and putting in the explicit form (11) for $f(r)$
$$\displaylines{W={c^4\over G}\,\int_0^\infty\,
{{Q^2\,e^{-Q^2/2Mr}}\over{8aMr^4}}\,\bigg[(4aMr^2+aQ^2r)\hfill\cr\hfill-(Q^2r^2-4a^2Mr+a^2Q^2)\,
\arctan\left({a\over r}\right)\bigg]\,dr\qquad (26) \cr}$$\ad This works out
simply to
$$W={c^4\over G}\,M=mc^2\eqn$$\ad implying again that the electron rest
mass is purely electromagnetic.

For distances $r\gg a,\,\,r_0$, the dominant components of the electric and
magnetic fields closely approximate a rotating ring of charge with radius $a$,
as previously shown in the analysis of Israel\cite{18} and of Pekeris and
Frankowski\cite{19}:

$$E_r\approx {e\over r^2}-{3e a^2 \cos^2\theta\over r^4} , \qquad
E_\theta\approx- {2e a^2\cos\theta\sin\theta \over r^4}$$
$$B_r \approx {2\mu\over r^3}\cos\theta,
\qquad B_\theta \approx {\mu\over r^3}\, \sin\theta \eqn$$\ad neglecting
terms of higher order in $a/r$ and $r_0/r$. We have used the following
conversions from geometrized units:
${\cal M}=Qa=QS/M=\sqrt{G}\mu/c^2,\quad Q=\sqrt{G}e/c^2$ and $S=Gs/c^3$.
The magnetic moment is given by
$$\mu = {e\hbar \over{2mc}} \eqn$$\ad for spin angular momentum
$s=\hbar/2$.  Remarkably, this corresponds to a spin g-factor of 2, in
agreement with Dirac theory, as first noted by Carter\cite{20}.

One might imagine, in concept, the spherically-symmetrical polarization charge
distribution of Eq (3) set into rapid rotation, causing parallel current elements
to attract, thus compressing the distribution into a disc and ultimately into a
Kerr-Newman ring.

In summary, we have proposed a plausible resolution of the long-standing
paradox for the classical electron---how it can be a point charge yet have
finite electromagnetic self-energy\cite{21}. The suggested mechanism involves
warping of spacetime in ultrastrong fields, leading to Hawking polarization of
the vacuum. We note that this further connection between general relativity
and quantum mechanics might contribute to a realization of a unified theory
extensively discussed by many workers, notably Wheeler\cite{22} and
Sachs\cite{23}.

\end{document}